\documentclass[twocolumn]{emulateapj}

\usepackage{lscape}
\usepackage{graphicx}
\usepackage{epsfig}
\usepackage{amsmath}

\newcommand{\Msolar}{M$_{\odot}$}
\newcommand{\kms}{km s$^{-1}$}
\newcommand{\eoi}{$e/i$}
\newcommand{\Nsample}{1344}              
\newcommand{\Nspec}{5201}                
\newcommand{\Nstars}{1144}               
\newcommand{\PRV}{P$_{RV}$}

\newcommand{\NMEM}{360}               
\newcommand{\NVAR}{55}                
\newcommand{\SM}{305}      
\newcommand{\SN}{452}      
\newcommand{\BM}{39}       
\newcommand{\BN}{17}       
\newcommand{\BLM}{16}      
\newcommand{\BLN}{68}      
\newcommand{\BU}{16}       
\newcommand{\U}{231}       

\newcommand{\mRV}{-8.16 $\pm$ 0.05 \kms}
\newcommand{\magn}{13.0$\leq$V$\leq$16.5}         
\newcommand{\massrange}{1.6 - 0.8 \Msolar}    

\begin{document}

\title{WIYN Open Cluster Study. XXXVIII. Stellar Radial Velocities in the Young Open Cluster M35 (NGC 2168)}
\shorttitle{WOCS. RV Measurements in M35}

\author{Aaron M. Geller\footnote{Visiting Astronomer, Kitt Peak National Observatory, National Optical Astronomy Observatory, which is operated by the Association of Universities for Research in Astronomy (AURA) under cooperative agreement with the National Science Foundation.},~Robert D. Mathieu$^*$,~Ella K. Braden$^*$,~S\o ren Meibom$^{*,}$\footnote{Current address: Harvard-Smithsonian Center for Astrophysics, 60 Garden Street, Cambridge, MA 02138, USA}}
\affil{Department of Astronomy, University of Wisconsin - Madison, WI 53706, USA}

\and

\author{Imants Platais}
\affil{Department of Physics and Astronomy, The Johns Hopkins University, Baltimore, MD 21218, USA}

\and

\author{Christopher J. Dolan$^*$}
\affil{Department of Astronomy, University of Wisconsin - Madison, WI 53706, USA}
\shortauthors{Geller et al.}

\begin{abstract}  

\vspace{1em}

We present \Nspec~radial-velocity measurements of \Nstars~stars, as part of an 
ongoing study of the young (150 Myr) open cluster M35 (NGC 2168). 
We have observed M35 since 1997, using the Hydra Multi-Object 
Spectrograph on the WIYN 3.5m telescope.  Our stellar sample covers
main-sequence stars over a magnitude range of \magn~(\massrange) and extends spatially 
to a radius of 30 arcminutes 
(7 pc in projection at a distance of 805 pc or $\sim$4 core radii).  
Due to its youth, M35 provides a sample of late-type stars with a range 
of rotation periods.  Therefore, we analyze the
radial-velocity measurement precision as a function of the projected rotational
velocity.  For narrow-lined stars ($v\sin i \leq$ 10 \kms), the radial velocities have a 
precision of 0.5 \kms, which degrades to 1.0 \kms~for stars with $v\sin i = 50$ \kms.
The radial-velocity distribution shows a well-defined cluster peak with a central 
velocity of \mRV, permitting a clean separation of the cluster and field stars.
For stars with $\geq$3 measurements, we derive radial-velocity 
membership probabilities and identify radial-velocity variables, finding \NMEM~cluster 
members, \NVAR~of which show significant radial-velocity variability.  Using these cluster 
members, we construct a color-magnitude diagram for our stellar sample cleaned 
of field star contamination.  We also compare the spatial distribution of 
the single and binary cluster members, finding no evidence for mass 
segregation in our stellar sample.  
Accounting for measurement precision, we place an 
upper limit on the radial-velocity dispersion of the cluster of $0.81 \pm 0.08$ \kms.  
After correction for undetected binaries, we derive a true radial-velocity dispersion of 
$0.65 \pm 0.10$ \kms.

\end{abstract}


\section{Introduction}

Young open clusters are laboratories for the direct study of the near-primordial characteristics of 
stellar populations.  Their properties, and particularly those of the binary systems, offer unique insights 
into how stars are born and provide essential guidance for $N$-body studies of star clusters.
Indeed, with sophisticated $N$-body simulations now able to model real open clusters
\citep[e.g.,][]{hurley:05}, knowledge of the correct initial conditions are all the 
more important.
In particular, the initial binary population has a vast impact on the dynamical evolution of the cluster,
and the characteristics of the initial binary population will affect the 
overall frequency, formation rate and formation mechanisms of anomalous stars, like blue 
stragglers, as interactions with binaries are thought to be catalysts for the formation of these
exotic objects \citep{hurley:05,knigge:09}.
As a rich open cluster with an age of $\sim$150 Myr, M35 is a prime
cluster to define these hitherto poorly known initial conditions for the 
binary population required for any open cluster simulation.  

M35 is a fundamental cluster in the WIYN Open Cluster Study \citep[WOCS;][]{mathieu:00},
and as such has a strong base of astrometric and photometric observations from both
the WOCS collaboration and others. The cluster is centered at 
$\alpha=6^{\rm h}09^{\rm m}07\fs5$ and
$\delta=+24\arcdeg20\arcmin28\arcsec$ (J2000), towards the Galactic anticenter.
Numerous photometric studies have identified the 
rich main-sequence population \citep[e.g.,][]{kalirai:03,vonhippel:02,sung:99}.
WOCS CCD photometry places the cluster at a distance of 805 $\pm$ 40 pc, with an age 
of 150 $\pm$ 25 Myr, a metallicity of [Fe/H]=-0.18 $\pm$ 0.05 and a reddening of 
$E(B-V)$=0.20 $\pm$ 0.01 (C. Deliyannis, private communication). The most 
recent published parameters, from \citet{kalirai:03}, place the cluster at a distance
of 912$\substack{+70 \\ -65}$ pc ($(m-M)_0$ = 9.80 $\pm$ 0.16) with an age of 180 Myr,
adopting a $E(\bv)$=0.20 and [Fe/H] = -0.21. (See \citet{kalirai:03} for a thorough
review of previous photometry references and their derived cluster parameters). 
We note that these two recent studies used different isochrone families. 

There have been multiple proper-motion studies of the cluster 
\citep{ebbinghausen:42,cudworth:71,mcnamara:86a}, although none 
determine cluster membership for individual stars fainter than $V \approx$ 15.0.  Using 
proper motions, \citet{leonard:89} derive a cluster mass from 1600-3200 
\Msolar~within 3.75 pc.  Detailed observations have also been made in M35 to study
tidal evolution in binary stars \citep{meibom:05,meibom:06,meibom:07}, lithium 
abundances \citep{steinhauer:04,navascues:01a}, and white dwarfs 
\citep{reimers:88,kurtis:04,kurtis:06,kurtis:09}.

This is the first paper in a series studying the dynamical state of M35 through 
the use of radial-velocity (RV) measurements.
The data and results presented in this series will form the largest database of 
spectroscopic cluster membership and variability in M35 to date. 
In this paper, we present results from our ongoing radial-velocity study of 
the cluster, which we began in September 1997.  Our stellar sample includes solar-type 
main-sequence stars within the magnitude range of \magn, which corresponds to a mass 
range\footnote{\footnotesize 
This mass range is derived from a 180 Myr Padova isochrone \citep{marigo:08} 
using the distance, reddening and metallicity from \citet{kalirai:03}.} of \massrange.
The main-sequence turnoff is at $V\sim$9.5, $\sim$4 \Msolar.
In Section~\ref{obs}, we describe this stellar sample, observations and data reduction 
in detail.  We thoroughly investigate our RV measurement 
precision and the effect of stellar rotation in Section~\ref{rotate}.  Then in 
Section~\ref{sec:results} we derive RV membership probabilities, and use our 
study of the RV precision to identify RV variables, which we assume to be binaries 
or higher-order systems.  Within this mass range, we identify \NMEM~solar-type
main-sequence members; \SM~are single\footnote{\footnotesize In the following, we use
the term ``single'' to identify stars with no significant RV variation. 
Certainly, many of these stars are also binaries, although generally with longer 
periods and/or lower mass ratios ($q=m_2/m_1$) than the binaries identified in this study.  
When applicable, we have 
attempted to reduce this binary contamination amongst the single star sample by photometrically 
identifying objects as binaries that lie well above the single-star main-sequence 
(see Section~\ref{vvar}).} (non-RV-variable)
stars while \NVAR~show significant RV variability.  We then use these results to plot a color-magnitude
diagram (CMD) cleaned of field star contamination, to search for evidence of 
mass segregation and to study the cluster RV dispersion (Section~\ref{sec:disc}).
Finally, in Section~\ref{sec:summary}, we provide a brief summary.
In future papers, we will study the binary population of M35 in detail, providing observations
that will be used to directly constrain the initial binary population of open cluster 
simulations.

\section{Observations and Data Reduction}
\label{obs}

In the following section, we define our stellar sample, provide a detailed description of our 
observations and data reduction process, and discuss the completeness of our spectroscopic 
observations.

\begin{figure}[!ht]
\epsscale{1.0}
\plotone{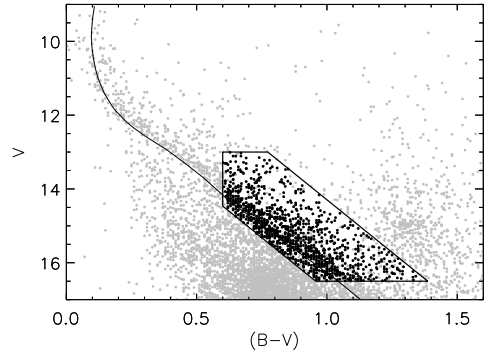}
\caption{\small Color-magnitude diagram for stars in the field of M35 highlighting the selected region used in this
survey.  We plot all stars in the field with the gray points to show the 
location of our selected sample relative to the full cluster.  Our stellar 
sample is bounded by the solid black lines.  Within this region, we plot 
observed stars in the solid black points.  Additionally, for reference we plot a 
180 Myr Padova isochrone \citep{marigo:08} using the distance, reddening and metallicity
 from \citet{kalirai:03}}
\label{cmd_bounds}
\end{figure}

\begin{figure*}[!ht]
\epsscale{1.0}
\plotone{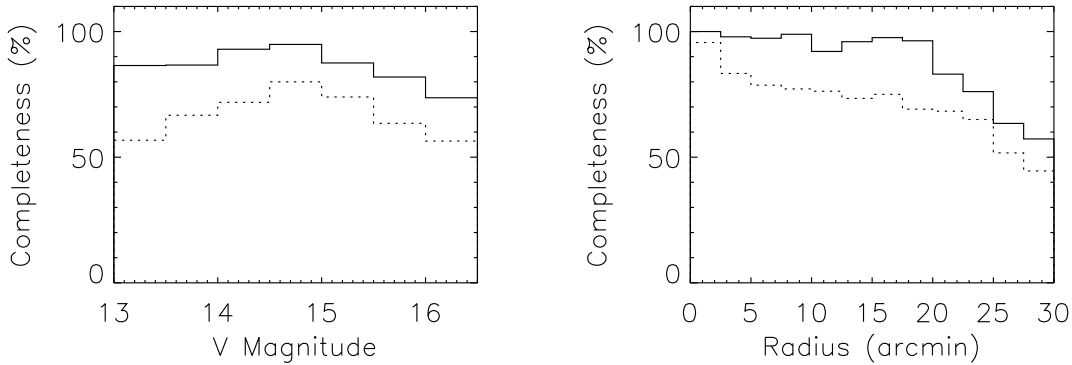}
\caption{\small Completeness of our observations as a function of $V$ magnitude 
(left) and projected radius (right).  We plot the completeness in stars 
observed $\geq$3 times with the dashed line, and stars observed $\geq$1 time 
with the solid line.}
\label{completeness}
\end{figure*}

\subsection{Photometric Target Selection}
\label{sub:sample}

Initially, we created our M35 target list from the stars in three wide-field 
CCD images centered on M35, taken by T. von Hippel with the Kitt Peak National 
Observatory (KPNO)  Burrell Schmidt telescope
on November 18 and 19, 1993.  These images have $V$ exposures of 4 s, 20 s and 180 s and $B$ 
exposures of 4 s, 25 s and 240s covering a 70$\arcmin$ $\times$ 70$\arcmin$ field.  We 
obtained $B$ and $V$ photometry with a limiting magnitude of $V=17$, denoted as 
source 1 in 
Table~\ref{RVtab}.  Additionally, we derive astrometry
from these plates, tied to the Tycho catalogue.

More recently, we added to our database the $BV$ photometry of Deliyannis 
(private communication), taken on the WIYN\footnote{The WIYN Observatory is a joint facility of the University of Wisconsin-Madison, Indiana University, Yale University, and the National Optical Astronomy Observatories.} 0.9m telescope 
with the S2KB 2K by 2K CCD. 
This photometry derives from a mosaic of five fields.  Each field has a
$20\arcmin \times 20\arcmin$ field-of-view, with one central field and four tiled around the center, 
for a total field-of-view of of $40\arcmin \times 40\arcmin$.  This 
photometry is denoted as source 2 in Table~\ref{RVtab}, and covers 74\% of the 
objects we have observed in this study.  We note that this photometry
is more precise than that of source 1.  The star-by-star difference in $V$ magnitudes for the 
two sources is roughly Gaussian with $\sigma$ = 0.06 mag.  However there is a tail
that extends beyond three times this sigma value.  Therefore we caution the reader
when using magnitudes from source 1.

We selected stars for the RV master list based on three constraints.
The faintest sources that can be observed efficiently at echelle 
resolution using the Hydra Multi-Object Spectrograph (MOS) on the 
WIYN 3.5m have $V$=16.5; this therefore sets our faint limit for 
observations.
Stars bluer than $(\bv)\sim0.6$ ($(\bv)_0\sim0.4$) do not provide precise
RV measurements due to rapid rotation and paucity of spectral lines; this 
therefore sets our blue limit for observations.
Finally, we perform a photometric selection of cluster member candidates,
shown as the outlined region\footnote{Specifically, we select stars with $0.6 < (\bv) < 1.5$, 
$13.0 < V < 16.5$ and between the lines defined by $5.7(\bv) + 8.6 < V < 5.7 (\bv)+11.0$.}  
 in Figure~\ref{cmd_bounds}.  This region
includes a wide swath above the main sequence so as to not select against binary stars 
\citep[e.g.][]{dabrowski:77}, yet also removes stars that are very likely cluster 
non-members.
This photometric selection allows for an efficient survey
of the cluster.  
Our sample extends radially to 30 arcminutes from the cluster center.
At a distance of 805 pc, this corresponds to the inner $\sim$7 pc 
of the cluster in projection.  Given the core radius derived by \citet{mathieu:83} of 
1.9 $\pm$ 0.1 pc, our sample is drawn from the inner $\sim$4 core radii.

Note that we lack $BV$ photometry for $\sim$11\% of the point sources 
found within 30 arcminutes from the cluster center in 2MASS.  For most of 
these objects, it is likely that there is a nearby or overlapping additional 
object which has prevented accurate $BV$ photometric measurements from either 
of our sources. These objects, by default, are not included in our stellar sample.  
In total, our stellar sample contains \Nsample~stars.

\subsection{Spectroscopic Observations}
\label{sec:observation}

Since September 1997, we have collected \Nspec~spectra of \Nstars~stars within 
this stellar sample as part of an ongoing observing program using the WIYN Hydra MOS.  
For the majority of these observations, we use Hydra's blue-sensitive 300 $\mu$m fibers, 
which project to a 3.1$\arcmin$ aperture on the sky.  We use the 316 lines mm$^{-1}$ echelle grating, 
isolating the 11th order with the X14 filter. The resulting spectra span a wavelength range 
of $\sim$25 nm, with a dispersion of 0.015 nm\ pixel$^{-1}$, centered on  512.5 nm.
We have also occasionally centered our observation on 637.5 nm using a very similar setup.
In this region, we use the same grating, but isolate the 9th order with the 
X18 filter.  These observations span a slightly larger wavelength range of $\sim$30 nm, 
and have a dispersion of 0.017 nm\ pixel$^{-1}$.
Due to a broken filter, observations taken after the 
spring of 2008 use different observing setups than discussed above; most are centered on 
560 nm and all use the echelle grating. 
We have not noticed any decrease in performance from the new wavelength range, but we caution the 
reader that we lack sufficient observations in these
setups to reliably determine our RV precision for these measurements.
During this same period certain upgrades were made to the spectrograph collimator\footnote{\footnotesize  http://www.astro.wisc.edu/$\sim$mab/research/bench\_upgrade/}.
All observed regions are rich in metal lines.  The typical velocity resolution
is 15 \kms.  In a two-hour integration,  the spectra have signal-to-noise (S/N) ratios 
ranging from $\sim$18 per resolution element for $V$=16.5 stars to
$\sim$100 per resolution element for $V$=13 stars.

We create fiber configurations (\textit{pointings}) for our observations 
using a similar method as \citet{geller:08}.  
Monte Carlo simulations show that we require at least three observations
over the course of a year in order to ensure 90\% confidence that a star is
either constant or variable in RV out to  binary periods of 1000 days
\citep[][Geller \& Mathieu, in preparation]{mathieu:83}.  Given three
observations with consistent RV measurements over a timespan
of at least a year and typically longer, we classify a given
star as single (strictly, non-RV variable) and finished, and move
it to the lowest priority. If a given star has three RV measurements
with a standard deviation $>$2.0 km s$^{-1}$ (four times our precision for 
narrow-lined stars; see Section~\ref{vvar}), we classify the star as RV variable and give
it the highest priority for observation on a schedule appropriate to its
timescale of variability.  This prioritization allows us to most efficiently
derive orbital solutions for our detected binaries.

We place our shortest-period binaries at the highest priority for
observations each night, followed by longer-period binaries to obtain
1-2 observations per run.  Below the confirmed binaries we place, in
the following order, ``candidate binaries'' (once-observed stars with
a RV measurement outside the cluster RV
distribution or stars with a few measurements that span only 1.5 - 2.5 km
s$^{-1}$), once observed and then twice observed non-RV-variable
likely members, twice observed non-RV-variable likely
non-members, unobserved stars, and finally, ``finished'' stars.
Within each group, we prioritize by distance from the cluster center,
giving those stars nearest to the center the highest priority.
A typical pointing will contain $\sim$70 fibers placed on individual 
stars in our sample and $\sim$10 sky fibers.

For a given pointing we obtain three consecutive exposures, each of 40 
minutes. In poor transparency or with a particularly bright sky, we restrict 
the targets to $V<15.0$ and shorten the integration time, generally to 20 minute
exposures. We obtain 
Thorium-Argon (ThAr), or occasionally Copper-Argon (CuAr), emission-lamp 
comparison spectra (300 s integrations) before and after each set of science 
integrations for wavelength calibration and to check for wavelength shifts 
during the observing sequence.  For each set of integrations we also obtain one 
flat-field image (200 s) of a white spot on the dome illuminated by 
incandescent lights.  Associating the flat-field images with the science 
integrations is particularly critical for calibrating throughput variations 
between the fibers in order to apply sky subtractions. In total, we have observed
106 distinct pointings in M35 over the roughly 11 years since our survey began.

\subsection{Data Reduction}
\label{sec:reduction}

For a thorough description of our data reduction process, see \citet{geller:08}.
In short, we perform a standard bias and flat-field correction to the 
images using the overscan strip and the flat-field images, respectively.  The 
flat-field spectra are used to trace each aperture in a given pointing and 
thereby extract the science spectra. Wavelength solutions derived from the 
emission-lamp spectra are applied, followed by sky-subtraction using sky fibers 
from each pointing.  The three sets of spectra (one set from each integration in 
a given configuration) are then combined via a median filter to remove 
cosmic ray signals and improve S/N.  
These reduced spectra are cross-correlated with a high S/N solar 
spectrum, obtained using a dusk sky exposure taken on the WIYN 3.5m with the 
same instrument setup as the given pointing. A Gaussian fit to the 
cross-correlation function (CCF) yields a RV and a full width at 
half maximum (FWHM, in \kms) for each stellar observation.  
The mean UT time is used to find and correct each RV measurement for the 
Earth's heliocentric velocity.  Finally we apply the unique fiber-to-fiber 
RV offsets derived by \citet{geller:08} for the WIYN-Hydra data to these 
RVs.  As in \citet{geller:08}, to ensure a sufficient quality of 
measurement, we incorporate into our database only those spectra with a 
CCF  peak height higher than 0.4.
Additionally, we examine the distribution of RVs for each individual star and
visually inspect any measurements that are outliers in the distribution.  
Occasionally we remove a measurement whose CCF, though having a peak height above 
0.4, clearly provides a spurious measurement (e.g., inadequate sky subtraction).

\subsection{Completeness of Spectroscopic Observations}
\label{sub:comp}

We have at least one observation for \Nstars~of the \Nsample~stars in our stellar sample,
for a completeness of 85\% across our entire sample.  60\% of the stars in our 
stellar sample have sufficient observations for their RVs to be considered 
final (813/\Nsample).  For these stars, we either have $\geq$3 RV 
measurements that show no variation, or, if we do see RV variability, 
we have found a binary orbital solution. (These 813 stars comprise the
SM, SN, BM and BN classes; see Section~\ref{sub:membership}).  
Of those stars not finalized, \U~have only one or two observations,
and another 100 stars are variable but do not yet have definitive orbital 
solutions.

In Figure~\ref{completeness}, we show the completeness of our observations as
functions of $V$ magnitude (left) and projected radius (right).  
We plot the completeness in stars 
observed $\geq$3 times with the dashed line and stars observed $\geq$1 time 
with the solid line.  Our prioritization of stars by distance from the 
cluster center is evident by our decreasing completeness with cluster 
radius.  The decreasing completeness towards fainter stars reflects 
the need for dark skies with minimal sky contamination in order to obtain
sufficient S/N in our spectra to derive reliable RVs for faint stars.
There are 37 stars with $V < 15$ in our stellar sample that do not have 
RV measurements, one of which is a proper-motion member. 
15 were observed but did not yield reliable RVs, mostly due to rapid rotation.
22 were not observed, 15 of which are farther than 20 arcminutes in radius from the cluster center.  

The difference in completeness between bright stars observed $\geq$1 and $\geq$3 times
is also a result of an increasing population of rapidly rotating stars towards bluer $(\bv)$ color.
For many of these stars, we have multiple observations of which only a few, and sometimes one, 
exceed this cutoff value of CCF peak height $>$0.4 and therefore are included in our database.  
For purposes of future research we also include seven rapid rotators in Table~\ref{RVtab} for 
which we have been unable to derive RVs from our spectra.

\section{Effects of Stellar Rotation on Measurement Precision}
\label{rotate}

\subsection{Observed Rotation}
\label{sub:rotobs}

Because of its youth, M35 provides a sample of late-type stars with a range of 
rotational periods \citep{meibom:09}; some of these stars have 
projected rotational velocities that exceed our spectral resolution. As such, 
the cluster presents an opportunity to explore empirically the dependence of 
our measurement precision on increasing $v\sin i$, where $i$ 
is the inclination angle of the stellar rotation axis to our line of sight.

\begin{figure}[!ht]
\plotone{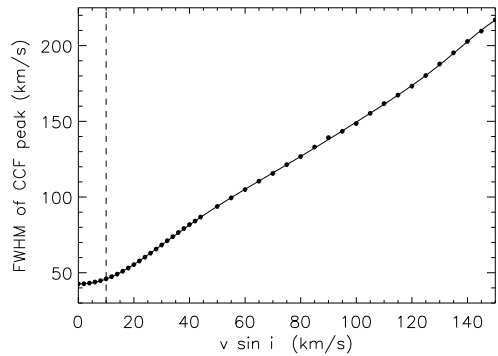}
\caption{\small FWHM as a function of $v\sin i$ for observations in the 512.5 nm region.  
FWHM values are measured from the CCF peaks derived from a series of artificially broadened 
templates, of known $v\sin i$, correlated against the original narrow-lined spectrum.
We also show a polynomial fit to the data, which we then use to derive $v\sin i$ values 
for observed stars in M35.  Additionally we plot a dashed line at $v\sin i$ = 10 \kms, 
below which the curve flattens out due to our spectral resolution. We impose a floor in 
$v\sin i$ at this value as we are unable to reliably measure slower rotation. }
\label{ftovfig}
\end{figure}

\begin{figure*}[!ht]
\epsscale{1.0}
\plottwo{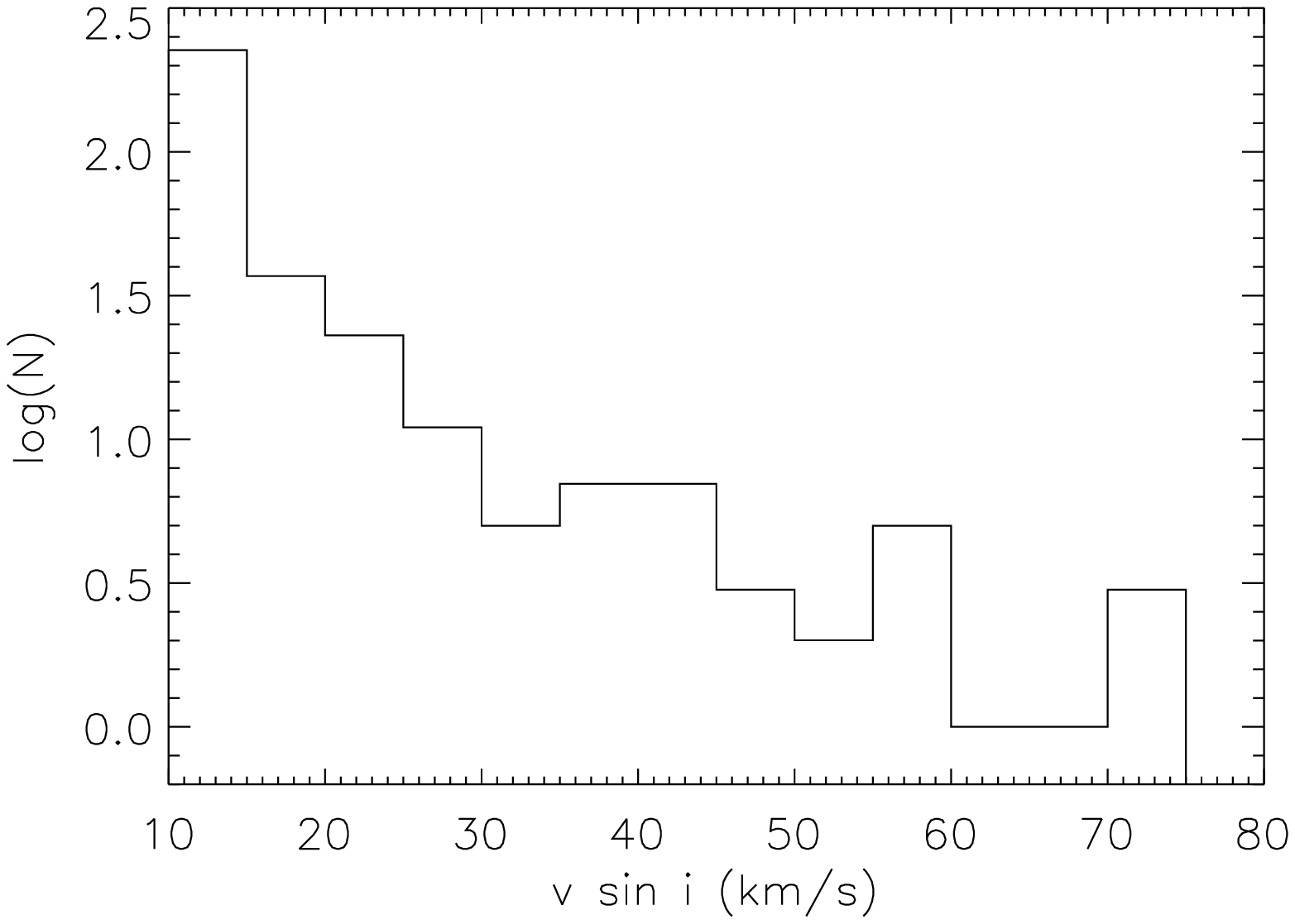}{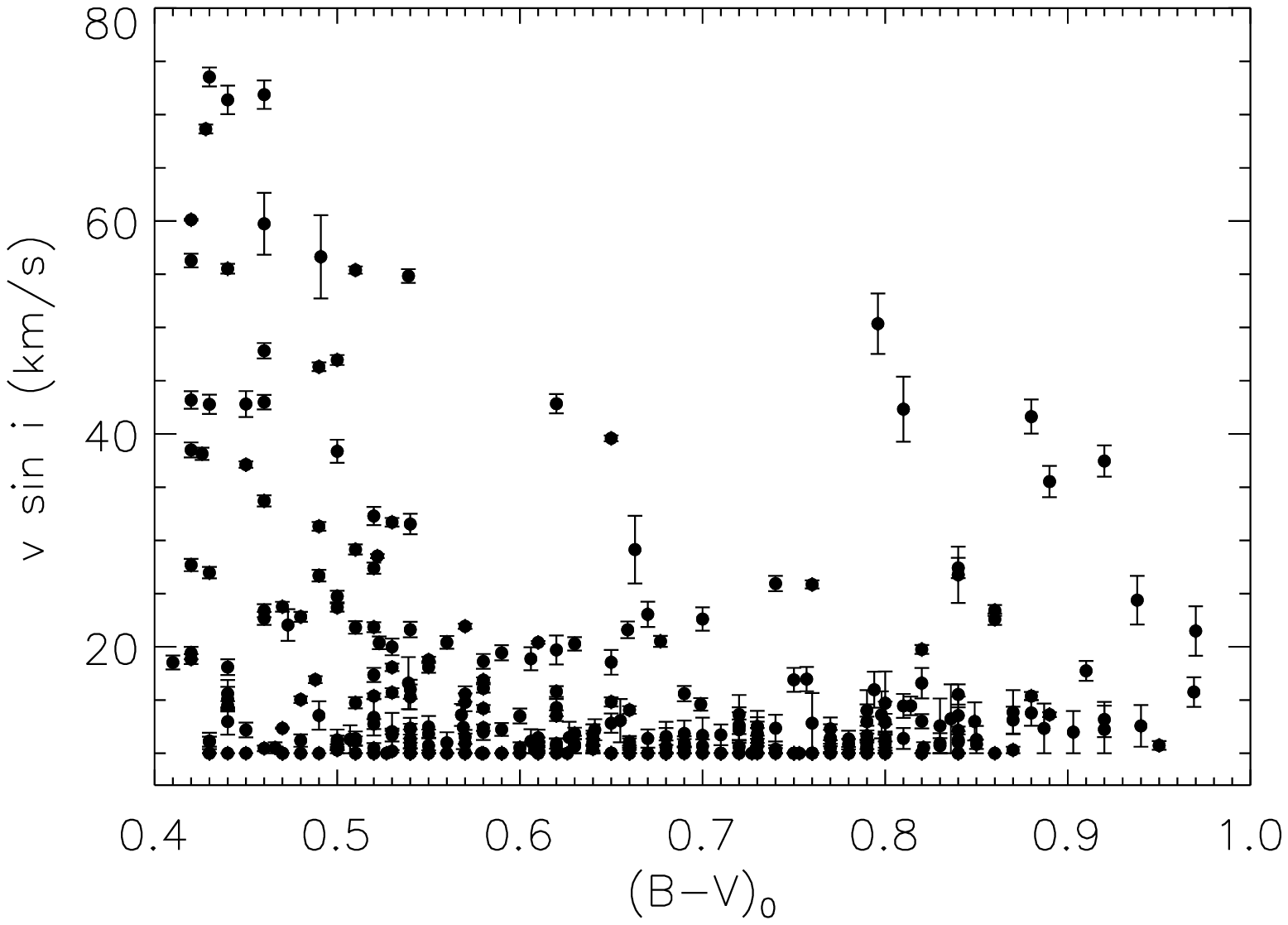}
\caption{\small Histogram of $v\sin i$ measurements (left) and $v\sin i$ as a function of 
$(\bv)_0$ (right) for the cluster members of M35.  
We have removed double-lined binaries and any binaries with known periods less than 10.2 days, the circularization
period in M35 \citep{meibom:05}.  We only show stars with mean $v\sin i$ values derived
from $\geq$3 observations within the 512.5 nm region.
Notice that the stars with the largest rotation are generally 
also the bluest stars in our sample.}
\label{vsini_BV}
\end{figure*}

The measured FWHM of the CCF for a given star is directly 
related to the $v\sin i$ \citep{rhode:01}.  
Thus in order to derive a $v\sin i$ value, we first measure the FWHM of the 
CCF peak. 
To do so, we fit a Gaussian function to the 
peak, forcing the baseline of the Gaussian to start at the background level of 
the CCF. Specifically, we subtract from the CCF a polynomial fit to this background
level, and then fit the Gaussian to the subsequent ``continuum subtracted'' CCF.
We only use spectra from the 512.5 nm region to measure the FWHM, as the 
FWHM is dependent on the setup (i.e., the dispersion, etc.), and most of 
our observations were taken in the 512.5 nm region.
We then use a similar technique as \citet{rhode:01}, to convert this FWHM to 
a $v\sin i$.  We create a series of artificially broadened templates by 
convolving our standard solar template with a series of theoretical rotation profiles of
specific $v\sin i$ values.  We then cross correlate this series of broadened templates 
with the original narrow-lined template and measure the FWHM of the CCF peak as described
above.  In Figure~\ref{ftovfig}
we show the results of this analysis along with a polynomial fit to the data.
We use this curve to derive $v\sin i$ values for all observations of stars in M35 in 
the 512.5 nm region.  We then take the mean $v\sin i$ for each star,  using only
our highest quality (CCF peak height $>$0.4) spectra, and provide these values in 
Table~\ref{RVtab}.
We are unable to reliably measure $v\sin i$ values below 10 
\kms, due to the spectral resolution; we therefore impose a floor to the $v\sin i$ 
at this value.

The median FWHM value that we observe is 46.1 \kms~which corresponds to $v\sin i$ = 10.3 \kms.
Excluding stars rotating slower than 10 \kms, we find a precision of 1.4 \kms~for 
individual $v\sin i$ values of $\leq$25 \kms, which increases to 1.6 \kms~for $v\sin i >$25 \kms.
These precision values were derived in the same manner as for our RV precision, with a fit to 
a $\chi^2$ function; see Section~\ref{sub:precision} and \citet{geller:08}.
Where possible, we derive a mean $v\sin i$ for a given star from multiple, generally $\geq$3,
observations within the 512.5 nm region.
We have compared our $v\sin i$ measurements to the rotation 
periods from \citet{meibom:09} 
for stars observed in both studies, and find the $v\sin i$ and rotation periods to be consistent.

In the left panel of Figure~\ref{vsini_BV}, we plot a histogram of the mean
$v\sin i$ measurements for M35 cluster members. (See Section~\ref{sub:membership} for
our membership criteria.)
In this and the other panel, we have excluded any binaries with periods known to be 
less than the circularization period in M35 of 10.2 days \citep{meibom:05}, as the rotation 
of the stars in these binaries have likely been affected by tidal processes.  
We have also removed any stars that appear
to be in double-lined binaries, as the spectral lines in many of these observations are 
broadened due to the secondary spectrum at similar, though slightly offset, RV.  
In the right panel of Figure~\ref{vsini_BV}, we plot the mean $v\sin i$ as a function of 
$(\bv)_0$ for M35 cluster members.  We see a clear trend of increasing 
rotation towards bluer stars, as has also been observed in other young open clusters
and the field
(e.g., field, Hyades, Pleiades, \citealt{kraft:67}; Pleiades, \citealt{soderblom:93}; 
Blanco 1, \citealt{mermilliod:08}; IC 2391, \citealt{platais:07}).

\subsection{Radial-Velocity Precision}
\label{sub:precision}

We determine the RV measurement precision following \citet{geller:08}, where a 
$\chi^2$ distribution is fit to the distribution of the standard deviations 
of the first three RV measurements for each star in an ensemble of 
stars. Here we do this operation on samples of stars with differing 
$v\sin i$. Specifically, we consider stars with $v\sin i$ of 
$\leq$10 \kms, 10 - 20 \kms~and 20 - 80 \kms. The bin sizes were chosen 
arbitrarily in order to provide sufficiently large samples.
The first bin contains all narrow-lined
stars for which we have imposed a floor to the $v\sin i$ (see Section~\ref{sub:rotobs});
these stars have line widths characteristic of the auto-correlation 
of our spectral resolution. The remaining bins contain stars with  
line widths increased by stellar rotation.

A detailed study of the RV measurement precision of our 
observation and data-reduction pipeline has been done by \citet{geller:08} for 
late-type stars in the old open cluster NGC 188. For the narrow-lined stars 
in NGC 188 they find a single-measurement precision of 0.4 \kms.
This precision is also a function of the S/N of the spectrum, as shown in 
\citet{geller:08} by the degrading precision with increasing $V$ magnitude
as well as decreasing CCF peak height.  The largest S/N effect seen for 
narrow-lined stars in NGC 188 is to degrade the precision by 0.25 \kms. 
The effect of rotation is larger than this amount.
Here, we derive a relationship between 
the measurement precision and $v\sin i$ and use this relationship in our analysis 
throughout this paper.  

\begin{figure}[!ht]
\plotone{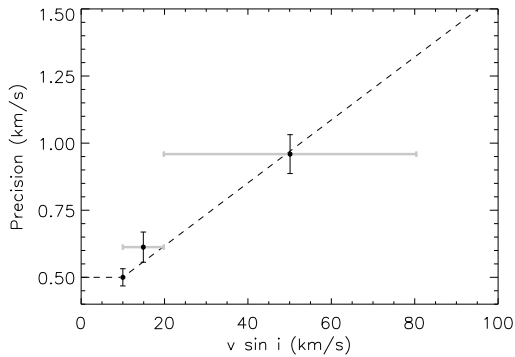}
\caption{\small RV measurement precision as a function of the average $v\sin i$
(in \kms) for single lined stars with $\geq$3 observations. The bins are $v\sin i$
of  $\leq$10 \kms, 10 - 20 \kms and 20 - 80 \kms, chosen 
to provide sufficiently large samples.  The gray horizontal bars indicate the 
bin sizes for each point.  The black vertical error bars show the one sigma 
errors on the precision fit values.  The dotted line shows the fit to these 
data, and provided in Equation~\ref{prec:eq}; we impose a floor to our precision
at 0.5 \kms.}
\label{vsini}
\end{figure}

In Figure~\ref{vsini} we show the RV precision as a function of $v\sin i$ in M35
for observations taken in the 512.5 nm region.  The narrow-lined stars 
have a RV precision of 0.5 \kms, similar to that found for the narrow-lined stars in 
NGC 188 observed with this same setup. As 
expected, the value of the measurement precision increases with increasing line 
width. For the most rapidly rotating stars ($v\sin i >$ 50 \kms), the measurement 
precision degrades to $\sim$1.0 \kms.  We fit a linear relationship to the points 
in Figure \ref{vsini}, shown as the dashed line:
\begin{equation}
\label{prec:eq}
\sigma_i=0.38 + 0.012 (~v\sin i~)~~\mathrm{km~s^{-1}},
\end{equation}   
where $\sigma_i$ is our precision.  
We use this equation with the 
mean measured $v\sin i$ for a given star to calculate the single-measurement RV
precision for that star.  
We adopt a floor to our precision at 0.5 \kms, as found 
for our narrow-lined stars, and shown by the break in the dashed line in Figure~\ref{vsini}.  

We lack sufficient observations to perform this same 
analysis using observations in the 637.5 nm region or for observations taken after 
the spring of 2008 (see Section~\ref{sec:observation}).  Therefore, for the 129 stars that 
do not have any observations in the 512.5 nm region ($\sim$11\% of our observed stars), 
we visually inspect the spectra and CCFs.  For narrow-lined stars, we set the precision to 
0.5 \kms, and for rotating stars we set the precision to 1.0 \kms.  
We can then use this RV precision value for a given star to determine
whether our observations for this star are constant or variable in velocity
(see Section~\ref{vvar}).
We note that only 13 of these stars have sufficient observations for their RVs to be considered 
final, and only 2 are probable members.

\section{Results}
\label{sec:results}

The full M35 database is available with the electronic version of this paper;
here we show a sample of our results in Table~\ref{RVtab}.  The first 
column in Table~\ref{RVtab} contains the WOCS identification number ($ID_W$).  These
numbers are defined in the same manner as in \citet{hole:09}, with the
cluster center set at  $\alpha=6^{\rm h}9^{\rm m}7\fs5$ and 
$\delta=+24\arcdeg20\arcmin28\arcsec$ (J2000). Next we give the 
corresponding IDs from \citet{meibom:09}, \citet{mcnamara:86a} and \citet{cudworth:71} ($ID_M$, $ID_{Mc}$ and $ID_C$). 
The next few 
columns provide the right ascension ($RA$), declination ($DEC$), the $BV$ photometry
and the source number ($S$) for this photometry (see Section~\ref{sub:sample}).  Next,
we show the number of RV measurements ($N$) and the mean and standard error of the RV 
measurements.
For stars with only one RV measurements, we show the single-measurement RV precision instead 
of the standard error.  Next we provide this single-measurement RV precision ($\sigma_i$, derived
using equation~\ref{prec:eq}),
the mean and standard error of the $v\sin i$ measurements\footnote{For double-lined binaries and 
stars with no observation in the 512.5 nm region, we do not derive a $v\sin i$ value.  
For stars with only one measurement in 
the 512.5 nm region, we convert the 1-sigma error on the FWHM (derived from the 
Gaussian fit to the CCF peak) to an error on the $v\sin i$ using the fit shown in Figure~\ref{ftovfig}.
As this relationship is not linear, we provide the mean of the derived upper and lower errors on 
$v\sin i$.},
the \eoi~value (see Section~\ref{vvar}), the calculated RV membership probability 
(\PRV, see Section~\ref{sub:membership}), the proper-motion membership 
probability from \citet{mcnamara:86a} ($P_{PM1}$) and \citet{cudworth:71} ($P_{PM2}$), where available, 
and then, the classification of the object (see Section~\ref{sub:class}). 
For RV-variable stars with orbital solutions, we present the center-of-mass 
($\gamma$) RV with the derived error in place of the mean RV and its standard error, and 
add the comment SB1 or SB2 for single- and double-lined binaries, respectively.
Additionally, for binaries without orbital solutions that appear to be double-lined, we add the 
comment of SB2. Finally, for purposes of future research we include seven rapid rotators for 
which we have been unable to derive RVs from our spectra, and label them with the comment RR.

\subsection{Membership}
\label{sub:membership}

\begin{figure}[!ht]
\plotone{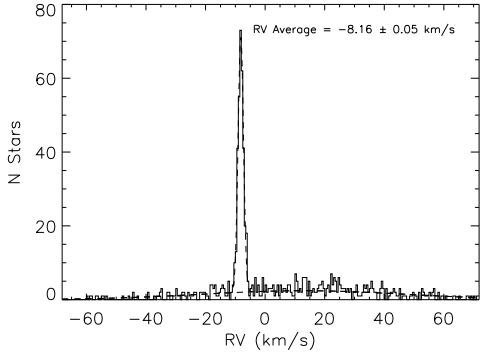}
\caption{\small RV histogram for stars in the field of M35. We include the mean RVs 
for stars observed $\geq$3 times with RV standard deviations $<$2 \kms~and 
the $\gamma$-RVs for binary stars with orbital solutions, excluding
RV variables whose $\gamma$-RVs are unknown.  The bin sizes are 0.5 \kms, 
equal to our RV precision for narrow-lined stars, as found in Section~\ref{rotate}.
The dashed lines show the simultaneous Gaussian fits to the cluster and field RV 
distributions.}
\label{vel_hist}
\end{figure}

The RV distribution of M35 is clearly distinguished from that of the field 
when we plot a histogram of the mean RVs for the observed stars in our stellar sample.
In Figure~\ref{vel_hist}, we show a histogram of the 
mean RVs for stars with $\geq$3 RV measurements whose standard deviations are 
$<$2 \kms, as well as the $\gamma$-RVs for binary stars with 
orbital solutions, thus excluding from the fit any RV variables
whose $\gamma$-RVs are unknown.  
The cluster shows a well-defined peak rising above the broad distribution of the field stars.
We simultaneously fit one-dimensional Gaussian functions, $F_{c}(v)$ and $F_{f}(v)$, to 
represent the cluster and field RV distributions, respectively, and then use these
fits to calculate RV membership probabilities for each individual star.
We compute the membership probability $P_{RV}(v)$ with the usual formula:
\begin{equation} \label{memeq}
P_{RV}(v) = \frac{F_{c}(v)}{F_{f}(v)+F_{c}(v)}  
\end{equation}
\citep{vasilevskis:58}. We plot these Gaussian fits in Figure~\ref{vel_hist} with the 
dashed lines, and show the fit parameters in Table~\ref{t:fit}.

For a given single star, we use the mean RV to compute the RV membership 
probability.  For a given binary star with an orbital solution, we compute the RV 
membership probability from the $\gamma$-RV. For RV-variable 
stars without orbital solutions, the $\gamma$-RVs are not known, and 
therefore we cannot calculate RV membership probabilities.  For these stars, we provide a
preliminary membership classification, described in Section~\ref{sub:class}.

\begin{deluxetable}{c r@{\hspace{0.5em}}c@{\hspace{0.5em}}l r@{\hspace{0.5em}}c@{\hspace{0.5em}}l}
\tablewidth{0pc}
\tablecolumns{3}
\tablecaption{Gaussian Fit Parameters For Cluster and Field RV Distributions
\label{t:fit}}
\tablehead{\colhead{} & \multicolumn{3}{c}{Cluster} & \multicolumn{3}{c}{Field}}
\startdata
Ampl. (Number) &                 69.0 &$\pm$& 2.0  & 2.4 &$\pm$& 0.4 \\
$\overline{RV}$ (km\ s$^{-1})$ & -8.17 &$\pm$& 0.05 & 13 &$\pm$& 4 \\
$\sigma$ (km\ s$^{-1})$ &         0.92 &$\pm$& 0.08 & 34 &$\pm$& 4 \\
\enddata
\end{deluxetable}

\begin{figure}[!ht]
\plotone{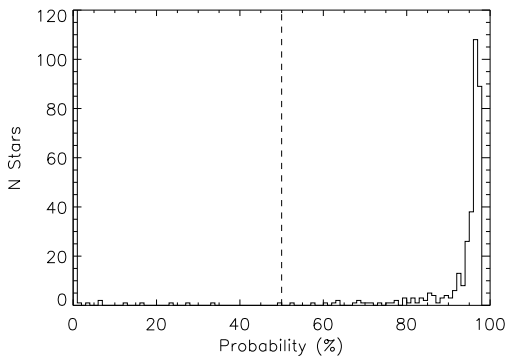}
\caption{\small Histogram of membership probabilities, \PRV, for stars observed $\geq$3 
times with RV standard deviations $<$2 \kms~and for binaries whose $\gamma$-RVs are known.  
For the single stars, we compute \PRV~using the 
mean observed RV; for binaries with orbital solutions, \PRV~is based on the $\gamma$-RV.
We show our membership cutoff of \PRV=50\% with the dashed line, above which
we classify a star as a cluster member.
Note that we do not show the full height of the bin at lowest membership probability for 
clarity.}
\label{prob_hist}
\end{figure}

In Figure~\ref{prob_hist}, we show the distribution of RV membership probabilities,
displaying a clean separation between the cluster members and field stars.  
In the following analysis, we use a probability cutoff of \PRV $\geq$ 50 \% to 
define our cluster member sample.  Using the 344 single cluster members and binary 
cluster members with orbital solutions, we find a mean cluster RV of \mRV.
From the area under the fit to the cluster and field distributions, 
we estimate a field contamination of 6\% within our cluster member sample (\PRV$\geq$50\%).  
Though this estimate is derived excluding the RV variables that do not have 
orbital solutions, the percent contamination should be valid for the cluster as a whole.

Our RV membership probabilities agree well with the proper-motion memberships of 
\citet{cudworth:71} and \citet{mcnamara:86a}.  We note that our stellar 
sample covers only the faintest portion of either proper-motion study.
There are 24 \citet{cudworth:71} proper-motion members within 
our observed stellar sample, of which we find 14 (58\%) to also have $\geq$50\% RV 
membership probabilities.  \citet{cudworth:71} note that for $V > 13$ they begin to find 
significant errors in their photometry and expect many field stars to 
contaminate their proper-motion member sample; this can likely explain 
the 10 discrepant stars.  There are 70 \citet{mcnamara:86a} proper-motion members 
within our observed stellar sample, of which we find 64 (91\%) to 
also have $\geq$50\% RV membership probabilities.  \citet{mcnamara:86a}
expects up to 15 field stars contaminating their cluster member sample from 
$13 < V < 15$, which can easily account for the 6 discrepant stars. 

We also note that NGC 2158 is only $\sim$28 arcminutes away from the center of M35, 
at $\alpha=6^{\rm h}07^{\rm m}25^{s}$ and $\delta=+24\arcdeg05\arcmin48\arcsec$ (J2000),
and thus is within the spatial region that we have surveyed.  \citet{scott:95} find a mean 
RV for NGC 2158 of $28 \pm 4$ \kms.  There are five stars within our sample that 
lie within the cluster radius of 2.5 arcminutes \citet{carraro:02} from the center of NGC 2158 and have 
RVs within three times the standard error  (12 \kms) of the mean RV : 
125044, 39017, 111050, 57037, 54048.  Two of these stars 
(125044 and 57037) have less than three observations; the remaining three have 
$\geq$3 observations and appear to be non-RV-variables.

\subsection{Radial-Velocity Variability}
\label{vvar}

RV-variable stars are distinguishable by the larger standard
deviations of their RV measurements.  Here, we assume that such velocity
variability is the result of a binary companion, or perhaps multiple
companions.  Specifically, we consider a star to be a RV variable
if the ratio of the standard deviation of its RV measurements to the
single-measurement RV precision\footnote{\footnotesize We use the same
nomenclature of ``\eoi'' as in \citet{geller:08}, though in other sections, for clarity, we have labeled 
the precision as $\sigma_i$, so as not to confuse the precision with an inclination angle.}
(\eoi) for that star is greater than four 
\citep{geller:08}.  We provide the \eoi~value for each single-lined star 
in Table~\ref{RVtab};  we label double-lined systems as 
RV variables directly, and include the comment of SB2 in 
Table~\ref{RVtab}.  

\begin{figure}[!ht]
\epsscale{1.0}
\plotone{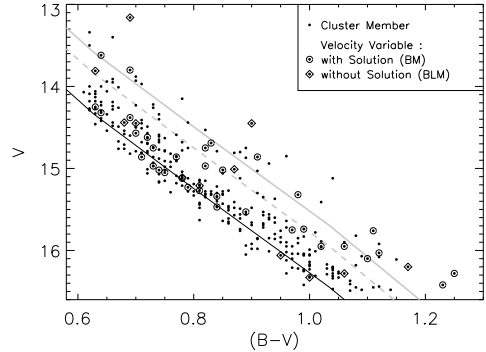}
\caption{\small Color-magnitude diagram of M35 including only cluster members (\PRV$\geq$50\%)
with photometry from WIYN 0.9m (source 2). We plot the RV variables with orbital solutions with 
circles and without orbital solutions with diamonds.  We show the 180 Myr Padova 
isochrone as the black line.  The solid gray line shows where binaries with mass ratios of 
1.0 lie on the CMD, and the dashed gray line shows the deviation from the isochrone of twice the 
photometric error.}
\label{CMDmem}
\end{figure}

Monte Carlo analysis has shown that, for similar observations of 
solar-type stars in NGC 188, Geller \& Mathieu (in preparation) can detect 
the majority of binaries with periods less than 10$^4$ days and
a negligible fraction of longer-period binaries.
Though the slightly poorer precision for the M35 data 
will effect the specific completeness numbers, we can assume 
a similarly high completeness in detected binaries with periods less than 10$^4$ days and
a corresponding drop in completeness for longer-period binaries.
Some of the undetected systems are evident from 
their separation from the main sequence (see Figure~\ref{CMDmem}).

We have currently identified \NVAR~RV-variable members of M35, and have derived 
orbital solutions for 71\% (\BM/\NVAR) of this sample.  
In following papers we will provide the orbital solutions for these 
systems, including all derived parameters.  We will then perform a detailed 
analysis of the distributions of these orbital parameters as well as the
binary frequency of the cluster.

\vspace{2em}

\subsection{Membership Classification of Radial-Velocity Variable Stars}
\label{sub:class}

We follow the same classification system as \citet{geller:08} and 
\citet{hole:09} in order to provide a qualitative guide to a given star's
membership and variability, in addition to the calculated RV memberships and 
\eoi~values.  We provide these classifications for all observed stars, while 
the memberships and \eoi~values are only provided for a subset of appropriate
stars.

\begin{deluxetable}{cc}
\tablewidth{0pt}
\tablecaption{Number of Stars or Star Systems Within Each Membership Class}
\tablehead{
\colhead{Class} & \colhead{Number} }
\startdata
SM & \SM \\
SN & \SN \\
BM & \BM \\
BN & \BN \\
BLM & \BLM \\
BU &  \BU \\
BLN & \BLN \\
U &  \U \\
\enddata
\label{t:class}
\end{deluxetable}

For stars with \eoi$<$4, we classify those with \PRV$\geq$50\% as single 
members (SM), and those with \PRV$<$50\% as single non-members (SN).  
If a star has \eoi$\geq$4 and enough measurements from which we are able to 
derive an orbital solution, we use the $\gamma$-RV to compute a secure 
RV membership.  For these binaries, we classify those with \PRV$\geq$50\% as
binary members (BM) and those with \PRV$<$50\% as binary non-members (BN).  
For RV variables without orbital solutions, we split our
classifications into three categories.  If the mean RV results in 
\PRV$\geq$50\%, we classify the system as a binary likely member (BLM).  If
the mean RV results in \PRV$<$50\% but the range of measured RVs
includes the cluster mean RV, we classify the system as a binary
with unknown membership (BU).  Finally, if the RV measurements for a
given star all lie either at a lower or higher RV than the cluster
distribution, we classify the system as a binary likely non-member
(BLN), since it is unlikely that any orbital solution could place the
binary within the cluster distribution.  We classify
stars with $<$3 RV measurements as unknown (U), as these stars do not
meet our minimum criterion for deriving RV memberships or \eoi~measurements.
In the following analysis, we include the SM, BM and BLM stars as cluster 
members.  Including these stars, we find \NMEM~total cluster members in our 
sample.  We list the number of stars within each class in Table~\ref{t:class}.

\section{Discussion}
\label{sec:disc}

In the following section, we present a CMD for M35 cleaned
of field star contamination (Section~\ref{sub:CMD}), compare the spatial distribution
of the single and binary members (Section~\ref{sub:spat}), and analyze the RV 
dispersion of the cluster (Section~\ref{sub:vdisp}).

\subsection{Color-Magnitude Diagram}
\label{sub:CMD}

In Figure~\ref{CMDmem}, we show the CMD for all RV cluster 
members in M35 from this study for which we have photometry from the WIYN 0.9m, as this 
set of photometry is of higher precision than that taken on the Burrell Schmidt (see 
Section~\ref{sub:sample}).  
We also plot a 180 Myr Padova isochrone using the 
cluster parameters derived by \citet{kalirai:03} in the black curve.  Binaries with orbital 
solutions are circled and RV variables without orbital solutions are marked by diamonds.

Additionally, we use the Padova isochrone to plot the location on the CMD of 
binaries with mass ratios $q = 1$, shown as the gray line.  
We note that there are a number of stars observed brighter and to the red of this 
line, some that we have not identified as RV variables. In this location on the CMD one 
would expect to find either higher-order systems or field stars.  
There are 33 RV members that lie above the $q = 1$ line;
22 are single and 11 show RV variability.  We expect a 6\% field star 
contamination within the cluster members sample (Section~\ref{sub:membership}).
If we include only the 309 cluster members that have photometry from the WIYN 0.9m (and are therefore
shown in Figure~\ref{CMDmem}), this results in 19 possible field stars; including our 
entire cluster member sample results in 22 possible field stars.
Therefore field star contamination cannot account for 
all of these sources, suggesting that a subset of these stars are indeed higher-order systems.
We also note that there are an additional 12 cluster members with photometry from
the Burrell Schmidt that lie above the $q = 1$ line, but recall that
this source of photometry is of poorer precision.

Finally, for use in Sections~\ref{sub:spat}, we follow a similar procedure
as \citet{montgomery:93} to attempt to photometrically identify binaries that lie far from the 
isochrone on the CMD.  
We derive the distance of each star from the main-sequence isochrone and fit
a Gaussian function representing the photometric error distribution to the distribution 
of these distances.  We notice a clear excess in the observed distribution from the Gaussian 
fit at 2$\sigma$,  shown as the dashed gray line in Figure~\ref{CMDmem}.  
We attribute this excess to photometric binaries.
A 1 \Msolar~star in M35 with the additional light from a companion
of mass-ratio $q = 0.78$ would lie on this line.
Therefore sources observed above this line are likely binaries with larger mass ratios ($q > 0.78$),
or very infrequently, field stars.  We observe 42 cluster members
above this line that show no significant RV variation (and therefore fall into the SM
class). Many of these are likely long-period binaries that 
are outside of our detection limits, as the hard-soft boundary for solar-type stars in M35 is 
$\sim10^5 - 10^6$ days, and we only detect binaries with $P \lesssim 10^4$ days (Geller \& Mathieu, in preparation).

\subsection{Spatial Distribution and Mass Segregation}
\label{sub:spat}

\begin{figure}[!ht]
\epsscale{1.0}
\plotone{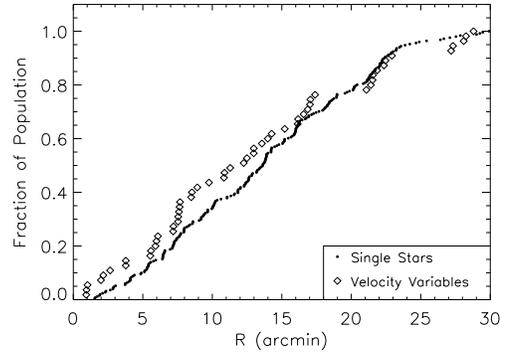}
\caption{\small Cumulative projected radial spatial distributions of the M35 single and 
binary cluster members.  We have excluded any stars from the single-star sample that 
are brighter and redder than the dashed grey line in Figure~\ref{CMDmem}, as these stars
are likely long-period binaries that are outside of our detection limits.  
We plot the single stars with the 
black points and the RV variables with the open diamonds.  We find no significant
evidence for central concentration of the RV-variable population.}
\epsscale{1.0}
\label{radial_cum}
\end{figure}

In Figure~\ref{radial_cum} we compare the cumulative projected radial distributions of the 
single and binary members of M35.  
We have attempted to reduce the contamination
from undetected binaries within our single-star sample by only including stars
with no detectable RV variation (SM) that are fainter and bluer than the dashed gray 
line in Figure~\ref{CMDmem}.
This conservative cut removes large-$q$ (i.e., high total mass) binaries that have periods 
longer than our detection limit. 
We have not applied any correction for the spatial bias found in our observations
(Section~\ref{sub:comp}), because this bias will be present in both
the single- and binary-star samples and should therefore not effect this analysis.
A Kolmogorov-Smirnov test shows no significant difference 
between these two populations with a value of 60\%.  We therefore conclude that the 
solar-type main-sequence binaries in M35 show no evidence for central concentration
as compared to the single stars.

\citet{mathieu:83} finds a half-mass relaxation time for the cluster of 150 Myr, 
comparable to the cluster age. This study of the radial spatial 
distributions for proper-motion-selected member stars in the $8.0 < V < 14.5$
($\sim$ 4.4 - 1.2 \Msolar) range revealed mass segregation only among stars more massive 
than 2 \Msolar. The degree of segregation lessens with decreasing mass, and is 
largely non-existent among solar-like stars. \citet{mcnamara:86b} found similar
results in their proper-motion selected sample, which covered stars down to
$V = 14.5$ ($\sim$ 1.2 \Msolar).  
We only include primary stars with masses of \massrange, and therefore 
most of our binaries have total masses that are lower than the 
higher-mass stars that have been shown to be mass segregated.
\citet{mathieu:83} found that M35
is fit well with a multi-mass King model. In such models the reduction in mass 
segregation for lower-mass systems derives from more severe tidal truncation of 
higher-dispersion velocity distributions in a cluster potential dominated by 
the solar-like stars. 

\subsection{Cluster Radial-Velocity Dispersion}
\label{sub:vdisp}

To determine the true RV dispersion of the cluster, we follow the procedure of \citet{geller:08}.
We first limit our sample to only include 
SM stars that have $v\sin i \leq$~10 \kms.  We limit the $v\sin i$ value 
to ensure that we only use the highest precision RV measurements for this analysis. 
These narrow-lined stars have a precision $\sigma_i$ = 0.5 \kms.  
We will discuss the effect of undetected binaries that likely remain within this sample 
in Section~\ref{sub:undetbin}.

Using this sample of 67 SM stars, we first derive the observed dispersion $\sigma_{obs}$ 
by taking the standard deviation of the mean RVs for each star,
and we find $\sigma_{obs} = 0.86 \pm 0.07$ \kms.
This observed dispersion is a function of our measurement precision and is also inflated by 
undetected binaries. 
Therefore, in order to derive the true RV dispersion, we must first account for the precision on these RV 
measurements.  
We derive\footnote{\footnotesize The use of Equations~\ref{sigcbeq}~and~\ref{xieq} is an
improvement over the procedure of \citet{geller:08} adapted from \citet{mcnamara:77}.  The uncertainty 
on $\sigma_{cb}$ also follows \citet{mcnamara:77}.}  
the ``combined RV dispersion'' $\sigma_{cb}$ from :

\begin{equation}
\label{sigcbeq}
\sigma_{cb}^2 = \sigma_{obs}^2 - \frac{1}{n}\sum_{i=1}^{n}\xi_i^2 .
\end{equation}

\noindent Here, $n = 67$ is the number of stars used in this analysis, and $\xi_i$ is the mean
error of the RV for the $i$th star defined as,

\begin{equation}
\label{xieq}
\xi_i = \left( \sum_{j=1}^m \frac{\left( RV_j - \overline{RV_i} \right) ^2}{m(m-1)} \right)^{1/2}
\end{equation}

\noindent where $RV_j$ is one of the $m$ number of RV measurements for a given star $i$, 
and $\overline{RV_i}$ is the mean RV for that star. We note that the second term in 
Equation~\ref{sigcbeq} is very nearly equal to $\sigma_i^2/3$, as 
we have 3 RVs for most of the stars in this sample, and these narrow-lined stars all have 
the same precision of $\sigma_i = 0.5$ \kms.  Following this procedure, we derive a 
combined dispersion of $\sigma_{cb} = 0.81 \pm 0.08$ \kms.  
The error on this combined dispersion is almost entirely due to the statistical error on $\sigma_{obs}$.

We find no significant difference in the combined RV dispersion of the SM
or BM stars.  For the BM stars, we use the $\gamma$-RVs in place of the 
mean RVs, and substitute the measurement precision for the standard deviation portion
in Equation~\ref{xieq}.
There is also no significant variation in the combined RV dispersion as a function of 
radius, although due to the small sample sizes our binned RV dispersion values 
have large uncertainties (of 0.1 - 0.15 \kms~for bins of 10 arcmin).

This combined RV dispersion is inflated by undetected binaries.  In the following section, 
we quantify this effect and apply the correction to derive the true RV dispersion of M35,  
an improvement on the procedure of \citet{geller:08}.


\subsubsection{Contribution from Undetected Binaries}
\label{sub:undetbin}

The combined RV dispersion defined in Equation~\ref{sigcbeq} is also described by, 

\begin{equation}
\sigma_{cb} = \sigma_c + \beta
\end{equation}

\noindent where $\sigma_c$ is the true RV dispersion of the cluster and $\beta$ represents the contribution from 
undetected binaries within our sample. Therefore, in order to derive the true RV dispersion of the cluster
we have performed a Monte Carlo analysis to determine this contribution from 
undetected binaries. 

We first create a set of simulated binaries 
with orbital parameters distributed according to the Galactic field solar-type binaries studied by \citet{duquennoy:91}.  
Specifically, these binaries have a log-normal period distribution
centered on log( $P$ [days] ) = 4.8 with $\sigma = 2.3$, and a Gaussian eccentricity 
distribution centered on $e = 0.3$. For binaries with periods below the circularization
period of 10.2 days \citep{meibom:05}, we set the eccentricity to zero.  
We use only solar-mass primary stars, and a distribution in secondary mass between 
0.08 - 1 \Msolar~described by a Gaussian centered on $M_2 = 0.23$ \Msolar~with $\sigma = 0.42$ 
\Msolar~\citep{kroupa:90}.  \citet{duquennoy:91} found this Gaussian to be the best fit to their 
solar-type field binaries, and this distribution is also consistent with that of \citet{goldberg:03} for 
their field binaries with primary masses $>$0.67 \Msolar.  
The orbital inclinations and phases of the binaries are chosen randomly.
We then generate three RVs for these simulated binaries distributed in 
time according to the actual distribution of our first three observations for 
stars in M35.  The majority of the SM stars in our sample have only three 
observations. To these RVs, we also add a random error generated from a Gaussian 
centered on zero and with $\sigma = 0.5$ \kms, the RV precision for narrow-lined stars in M35.
We also add a random velocity offset generated from a Gaussian centered on zero
with a standard deviation equal to an adopted one-dimensional RV dispersion.

\begin{figure}[!ht]
\epsscale{1.0}
\plotone{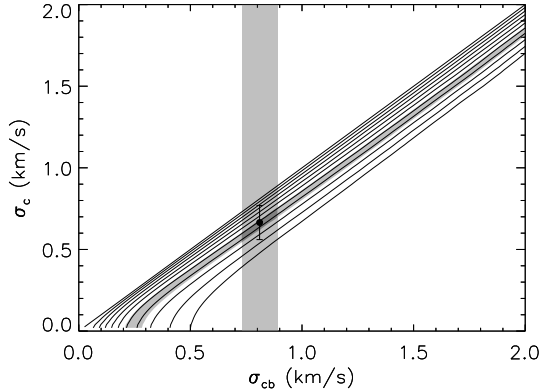}
\caption{\small The true cluster RV dispersion ($\sigma_c$) plotted against the 
combined RV dispersion ($\sigma_{cb}$) for a range of total binary frequencies. 
Each line corresponds to a different binary frequency in steps of 10\%, with 0\% at the 
far left and 100\% at the far right.  
With the vertical gray rectangle, we plot the region
included in the combined RV dispersion for M35 of $0.81 \pm 0.08$ \kms.  
The diagonal gray region covers the possible lines within our extrapolated true binary 
frequency in M35 of 
66\% $\pm$ 8\% 
(derived assuming the M35 binaries follow a 
\citet{duquennoy:91} period distribution).  Finally, we plot the resulting true
RV dispersion in M35 of 
0.65 $\pm$ 0.10 \kms~with 
the black point at the 
intersection of these two shaded regions.}
\epsscale{1.0}
\label{undbin}
\end{figure}

To this sample, we add a number of simulated single stars to produce a desired
binary frequency.  We generate three RVs for each single star from 
a Gaussian described by our precision.  To the mean RV for every single 
star we also add a random offset described by the assumed RV dispersion 
in the same manner as for the simulated binaries.
We then keep only those simulated binaries (and single stars) whose first three RVs
result in an \eoi~$<$~4, and whose mean RVs are within three 
standard deviations of the mean RV from a Gaussian fit to the simulated RV distribution.  This 
cutoff in standard deviation reflects our membership criterion of \PRV~$\geq$ 50\% for the M35 observations.
These binaries would be undetected within the SM sample.

We then follow the equations given above to derive $\beta$ for a range of binary 
frequencies and velocity dispersions, $\sigma_c$.  
In Figure~\ref{undbin}, we plot the true cluster RV dispersion ($\sigma_c$) as a function of the
combined RV dispersion ($\sigma_{cb}$) for a range of total binary frequencies, where 
each line corresponds to a different binary frequency between 0\% (far left) to 100\% (far right)
in steps of 10\%. 
We can then use the results shown in Figure~\ref{undbin} to derive the true RV dispersion for M35.
Furthermore, the results shown in this figure are also applicable to RV dispersion analyses
for other star clusters, provided that the binary population is consistent with the 
\citet{duquennoy:91} field binaries.

\subsubsection{True Radial-Velocity Dispersion}

To date, we have detected \NVAR~binaries in M35 out of \NMEM~cluster members.
If we assume a similar completeness as in Geller \& Mathieu (in preparation), for NGC 188,
then we can assume that we have detected 
63\%
of the binaries with periods 
less than 10$^4$ days (and a negligible fraction of binaries with longer periods).  
This correction results in a binary frequency of 
24\% $\pm$ 3\%
for $P < 10^4$ days.  This binary frequency is consistent with that of solar-type 
stars in the Galactic field \citet{duquennoy:91} out to the same period limit.
If we assume the M35 binaries follow a \citet{duquennoy:91} period
distribution, then our binary frequency for $P < 10^4$ days implies
a total binary frequency of 
66\% $\pm$ 8\%,
with the inclusion of wider binaries currently beyond our detection limits.
We then take this value for the total binary frequency and 
correct our combined RV dispersion for undetected binaries.  

In the filled gray areas in Figure~\ref{undbin} we show the regions defined by our M35 combined RV dispersion and 
the total binary frequency.
At the intersection, we plot the derived true RV dispersion in M35 of 
$\sigma_c = 0.65 \pm 0.10$ \kms.
Using a flat distribution in secondary mass (and mass ratio), as has been suggested by some studies 
\citep[e.g.,][]{mazeh:92,mazeh:03}, has a negligible effect on the 
derived true RV dispersion.
This true RV dispersion is consistent with the projected velocity dispersion of 
1.0 $\pm$ 0.15 \kms, derived by \citet{leonard:89} using the proper-motion data from \citet{mcnamara:86a}.

\section{Summary}
\label{sec:summary}

This is the first paper in a series studying the dynamical state of the 
young ($\sim$150 Myr) open cluster M35 (NGC 2168).   In this first paper, 
we present our RV observations and provide initial results from this survey.
Our stellar sample extends to 30 arcminutes
in radius from the cluster center (7 pc in projection at a distance of 805 pc or 
$\sim$4 core radii), and we have selected a region from a $V$, $(\bv)$ CMD 
(Figure~\ref{cmd_bounds}) which covers a mass range of \massrange.
We have used the WIYN 3.5m telescope with the Hydra MOS to obtain 
\Nspec~spectra of \Nstars~stars within this stellar sample.  From these 
spectra, we derive RV measurements with a precision of 0.5 \kms~for 
narrow-lined stars. The vast majority of the observed stars have multiple 
measurements, allowing determination of cluster membership and 
identification of spectroscopic binary stars.  We detect \NMEM~cluster 
members, \NVAR~of which show significant variability in their RV 
measurements.  Binary orbital solutions have been obtained for \BM~of these
RV variables, which we will present in detail in the next paper 
in this series. Observations of the rest of the RV variables 
and the remainder of our stellar sample are ongoing.
Table~\ref{RVtab} provides the first RV membership database for M35 and 
extends $\sim$1.5 magnitudes deeper than any previous membership catalogue.

Using the RV cluster members, we study the spatial distribution and velocity 
dispersion of the single and binary stars.
We find their spatial distributions to be indistinguishable.
This lack of central concentration for the binaries is 
consistent with earlier observational studies of stars in M35 as well as with 
a fully relaxed dynamical model for the cluster \citep{mathieu:83,mcnamara:86b}.  
In these studies, mass segregation is 
seen in higher-mass stars, but diminishes to being undetectable for stars in 
our observed mass range.  
After correcting for measurement precision, but not for binaries, we place an 
upper limit on the RV dispersion of the cluster of $0.81 \pm 0.08$ \kms.  
When we also correct for undetected binaries, we derive a true RV dispersion of 
$0.65 \pm 0.10$ \kms.

The WOCS group will continue our survey of M35 in order to derive RV memberships
for all stars in our stellar sample and obtain orbital solutions for all binaries
with periods less than a few thousand days, as well as some with longer periods.
In future papers, we will study the binary population of M35 in detail, providing 
all orbital solutions and analyzing the binary frequency and distributions
of orbital parameters.  These data will form essential constraints on the 
hitherto poorly known initial binary populations used in sophisticated $N$-body models of open clusters.

\acknowledgments
The authors would like to express their gratitude to the staff of the WIYN 
Observatory for their skillful and dedicated work that have allowed us to obtain 
these excellent spectra.  We thank Ata Sarajedini and Ted von Hippel for the 
acquisition of the Schmidt images, Vera Platais for work on the astrometry and 
photometry as well as John Bjorkman for early photometry work.  We also thank the many 
undergraduate and graduate students who have contributed late nights to obtain the spectra for 
this project.  This work was supported by NSF grant AST 0406615 and the 
Wisconsin Space Grant Consortium.

Facilities: \facility{WIYN 3.5m}

\bibliographystyle{apj}     
\bibliography{m35.rv.ms}

\tabletypesize{\tiny}
\clearpage
\begin{landscape}
\begin{deluxetable}{lllccccccccccccccccll}
\tablecaption{Radial-Velocity Data Table \label{RVtab}}
\tablewidth{0pt}
\tablehead{\colhead{$ID_W$} & \colhead{$ID_M$} & \colhead{$ID_{Mc}$} & \colhead{$ID_C$} & \colhead{$RA$} & \colhead{$DEC$} & \colhead{$V$} & \colhead{$(\bv)$} & \colhead{$S$} & \colhead{$N$} & \colhead{$\overline{RV}$} & \colhead{$RV_e$} & \colhead{$\sigma_i$} & \colhead{$\overline{v\sin i}$} & \colhead{$(v\sin i)_e$} & \colhead{$P_{RV}$} & \colhead{$P_{PM1}$} & \colhead{$P_{PM2}$} & \colhead{$e/i$} & \colhead{Class} & \colhead{Comment}}
\startdata
 96041     &     410 & \nodata & \nodata & 6:10:28.65 & 24:11:52.0 & 16.416 &  0.960 &  1 &   4 &   51.89 &    1.55 &    0.52 &     11.8 &      1.1 & \nodata & \nodata & \nodata &    5.93 &     BLN & \nodata \\
 36042     &     209 & \nodata & \nodata & 6:10:34.30 & 24:14:07.8 & 14.835 &  0.859 &  1 &   3 &   -8.74 &    0.55 &    0.64 &     21.6 &      0.8 &      96 & \nodata & \nodata &    1.49 &      SM & \nodata \\
 36045     &     209 & \nodata & \nodata & 6:10:43.69 & 24:16:08.9 & 14.497 &  0.824 &  1 &  17 &   -9.58 &    0.19 &    0.50 &     10.3 &      0.2 &      91 & \nodata & \nodata &   77.46 &      BM & SB1 \\
138057     &     366 & \nodata & \nodata & 6:10:50.20 & 24:04:50.7 & 16.368 &  1.165 &  1 &   1 &  -25.13 &    0.50 &    0.50 &  \nodata &  \nodata & \nodata & \nodata & \nodata & \nodata &       U & \nodata \\
 64052     &     312 & \nodata & \nodata & 6:10:43.70 & 24:07:00.8 & 15.884 &  1.023 &  1 &   4 &   -8.85 &    0.64 &    0.55 &     14.2 &      4.2 &      96 & \nodata & \nodata &    2.34 &      SM & \nodata \\
 15036     &     180 & \nodata &     731 & 6:10:15.70 & 24:11:31.7 & 13.450 &  0.690 &  2 &   1 &   57.98 &    0.50 &    0.50 &  \nodata &  \nodata & \nodata & \nodata &       0 & \nodata &       U & \nodata \\
 49051     &     227 & \nodata & \nodata & 6:10:51.32 & 24:11:10.6 & 15.086 &  0.890 &  1 &   4 &   88.83 &    0.34 &    0.50 &     10.0 &  \nodata &       0 & \nodata & \nodata &    1.35 &      SN & \nodata \\
 29047     &      87 & \nodata & \nodata & 6:10:44.59 & 24:13:44.3 & 14.948 &  0.895 &  1 &   4 &   56.95 &    0.28 &    0.50 &     10.0 &  \nodata &       0 & \nodata & \nodata &    1.10 &      SN & \nodata \\
 40032     & \nodata & \nodata & \nodata & 6:10:11.15 & 24:14:01.5 & 15.280 &  0.820 &  2 &   3 &   -7.72 &    0.32 &    0.55 &     14.3 &      0.8 &      96 & \nodata & \nodata &    1.02 &      SM & \nodata \\
 20037     &     193 & \nodata &     758 & 6:10:22.30 & 24:14:39.3 & 14.270 &  0.650 &  2 &   1 &    3.88 &    0.50 &    0.50 &  \nodata &  \nodata & \nodata & \nodata &       7 & \nodata &       U & \nodata \\
\enddata
\end{deluxetable}
The contents of each column are defined in Section 4.

\clearpage
\end{landscape}
\tabletypesize{\footnotesize}

\end{document}